\documentclass[useAMS,usenatbib,article]{mn2e}

\usepackage{aas_macros}
\usepackage{graphicx}
\usepackage{amssymb}
\usepackage{array}
\usepackage{multirow}
\usepackage{xcolor}
\title[Hydrodynamics of galaxy mergers with SMBHs]{Hydrodynamics of galaxy mergers with supermassive black holes: is there a last parsec problem~?}
\author[D.~Chapon, L.~Mayer and R.~Teyssier]{\parbox[t]{\textwidth}{Damien Chapon$^{1}$\thanks{E-mail: damien.chapon@cea.fr}, Lucio Mayer$^{2}$ and Romain Teyssier$^{1,2}$ }\vspace*{6pt}\\
$^{1}$CEA Saclay, DSM/IRFU/SAP, B\^atiment 709, F-91191 Gif-sur-Yvette Cedex, France\\
$^{2}$Institute for Theoretical Physics, University of Z\"urich, Winterthurestrasse 190, CH-8057 Z\"urich, Switzerland}
\begin{document}
\pagerange{\pageref{firstpage}--\pageref{lastpage}} \pubyear{2011}
\maketitle

\label{firstpage}

\newcommand{\Msun}{\textrm{M}_{\odot}}
\newcommand{\mattext}[1]{\textrm{\tiny #1}}
\newcommand{\Mach}{\mathcal{M}}
\newcommand{\Myr}{\;\textrm{Myr}}
\newcommand{\kms}{\;\textrm{km}/\textrm{s}}
\newcommand{\Hcc}{\;\textrm{H}/\textrm{cc}}
\newcommand{\fgas}{f^{\mattext{(gas)} }}
\newcommand{\tauOD}{\tau_{\mattext{od} }}
\newcommand{\csound}{c_{\mattext{s}} }
\newcommand{\Gammad}{\Gamma_{\mattext{d} }}
\newcommand{\Gammah}{\Gamma_{\mattext{h} }}
\newcommand{\RAMSES}{\texttt{RAMSES }}
\newcommand{\GASOLINE}{\texttt{GASOLINE }}
\newcommand{\comment}[1]{\begin{color}{purple}$\bullet$ \textbf{#1} $\bullet$\end{color}}

\begin{abstract}
We study the formation of a supermassive black hole (SMBH) binary and the shrinking of the separation of the two holes to sub-pc scales starting from a realistic major merger between two  gas-rich spiral galaxies with mass comparable to our Milky Way.
The simulations, carried out with the Adaptive Mesh Refinement (AMR) code {\ttfamily RAMSES}, are capable of resolving separations as small as 0.1 \textrm{pc}.
The collision of the two galaxies produces a gravo-turbulent rotating nuclear disk with mass ($\sim10^{9} \; \Msun$) and size ($\sim 60 \textrm{pc}$) in excellent agreement with previous SPH simulations with particle splitting that used a similar setup \citep{MayerSci2007} but were limited to separations of a few parsecs.
The AMR results confirm that the two black holes sink rapidly as a result of dynamical friction onto the gaseous background, reaching a separation of 1 \textrm{pc} in less than $10^{7} \; \textrm{yr}$.
We show that the dynamical friction wake is well resolved by our model and we find good agreement with analytical predictions of the drag force as a function of the Mach number.
Below 1 \textrm{pc}, black hole pairing slows down significantly, as the relative velocity between the sinking SMBH becomes highly subsonic and the mass contained within their orbit falls below the mass of the binary itself, rendering dynamical friction ineffective.
In this final stage, the black holes have not opened a gap as the gaseous background is highly pressurized in the center. Non-axisymmetric gas torques do not arise to restart sinking in absence of efficient dynamical friction, at variance with previous calculations using idealized equilibrium nuclear disk models.
We believe that the rather "hot" Equation-of-State we used to model the multiphase turbulent ISM in the nuclear region is playing an important role in preventing efficient SMBH sinking inside the central parsec.
We conclude with a discussion of the way forward to address sinking in gaseous backgrounds at sub-pc scales approaching the gravitational wave regime. 
\end{abstract}

\begin{keywords}
Galaxies: Mergers -- Galaxies: Structure -- Black Holes: Evolution -- Black Holes: Binaries -- Methods: Numerical -- Methods: Hydrodynamics
\end{keywords}

\section{Introduction}

The coalescence of two supermassive black holes (SMBHs), namely black holes in the mass range $10^5 - 10^9 M_{\odot}$, would produce the loudest gravitational wave signals in the Universe, and is the main target of planned low-frequency gravitational wave experiments, such as space-based laser interferometers and pulsar timing arrays \citep{Vecchio2004, Sesana2009}.
Since all massive galaxies are believed to host a central supermassive black hole, in the current cosmological paradigm in which galaxies grow via repeated mergers, mergers between such black holes could be common, especially at high redshift when the galaxy merger rate is higher.
However, the merger rate of SMBHs does not follow trivially from that of their host galaxies. Once the two galaxy cores have merged, leaving no distinct substructure at hundred of parsecs/kiloparsec scales, the two black holes will need to reduce their separation to less than $0.01$ pc before they can start to lose orbital energy efficiently via gravitational wave emission and eventually coalesce \citep*{BBR1980}.
Loss of orbital energy can occur via dynamical friction onto the stellar background \citep{MilosavljevicMerritt2001} or due to the gas drag \citep{Escala2004}. Both mechanisms are relevant since SMBHs are inferred to exist at the center of both gas-rich spirals and gas-poor ellipticals/S0s \citep*{VolonteriHaardtGultekin2008}.
The first one is known to become ineffective when the binary begins to harden; at this stage the 3-body interactions between the binary and individual stars deplete its loss cone, which can be refilled only via some orbital diffusion mechanism that brings fresh stellar material from other regions of the galaxy \citep{Berczik2005}.

In recent years it has been shown that SMBH decay in a gas dominated system is much more effective at forming rapidly a binary of SMBHs, in only a few million years following the major merger of two moderately gas-rich disk galaxies \citep{MayerSci2007, MayerKazantzidis2008}.
The reason is that the decay occurs in a much denser medium relative to the stellar dominated case, and because the black holes move with Mach numbers of order unity or slightly larger, the dynamical friction onto a gaseous medium is expected to be more effective than in a collisionless system \citep{Ostriker1999}.
Indeed the two black holes spiral down to parsec scales in a gaseous, dense nuclear disk a hundred pc in size formed by the dramatic gas inflow in the merger. Such nuclear disks have been found in high resolution multi-wavelength observations of nearby merger remnants \citep[][etc]{DownesSolomon1998, Davies2007}.
In minor mergers efficient black hole pairing at scales of tens of parsecs, namely even before a bound binary can form, requires high gas fractions ($> 30\%$) in both galaxies \citep{Callegari2009, Callegari2011} strengthening even further the crucial role of gas in the pairing and merging of SMBHs.
Other investigations starting when the two black holes form already a loosely bound pair in a nuclear disk have suggested that the decay can continue to sub-pc scales under certain conditions but have utilized pre-defined models of nuclear disks \citep{Escala2005, Dotti2006, Dotti2007, Cuadra2009} rather than starting from a realistic galaxy merger and appeal to tidal torques rather than dynamical friction when the binary begins to harden.
At the smallest scales, these models assumed very specific configurations such as that of two black holes already at less than $0.1$ pc separation embedded in a gap within the disk \citep{Cuadra2009}.
Gap opening implies the transition to a regime in which tidal torques from the surrounding disk become the dominant mode to extract energy and angular momentum from the binary.
However, the configuration of the host on which the two black holes are found at less than parsec scales is not really known because no computation exists that can reach such a stage starting from a realistic merger, and consequently the decay in such regime is not yet explored by three-dimensional simulations.

Until now, studies of binary black hole formation and shrinking in a gaseous environment have been based exclusively on SPH simulations, which might be unable to capture the turbulent nature of the flow in the disk.
Turbulence arises as a result of shocks during the final galaxy collision, and later in the nuclear disk as a result of its gravitational instability \citep{MayerSci2007, MayerKazantzidis2008}. Furthermore, dynamical friction, which is the central physical process involved, is numerically challenging to capture since it involves both the effect of the local overdensity trailing the black holes and that of the larger scale torques and tidal distortions generated by the surrounding mass distribution \citep{ColpiMayerGovernato1999}.

While TreeSPH codes such as \GASOLINE \citep{WSQGasoline2004} are typically well suited to address processes in the domain of self-gravitating systems, to which category dynamical friction belongs, particle noise and the difficulty to resolve gradients between media of different densities (such as the overdense wake and the surrounding background), may cast doubts on the quantitative results regarding the strength of dynamical friction.
More generally, confirmation of the results of SPH calculations on the effectiveness of the SMBH binary formation process in a gaseous environment is desirable. Due to the many scales involved, adaptive mesh refinement (AMR) simulations are ideally suited to the problem. In this paper, we used the \RAMSES code \citep{Teyssier2002} to follow both the galaxy merger at large scale and the binary SMBHs merger at small separation.
\RAMSES is based on a Particle-Mesh N-body integrator for dark matter and stars and a second-order unsplit Godunov scheme for the gas. Its shock capturing abilities and low intrinsic numerical viscosity (we used the HLLC Riemann solver and the MinMod slope limiter) make it ideal to address the problem.
It also features a multigrid Poisson solver that is both fast (of order N) and second-order accurate \citep{Guillet2011}, this being another desirable feature for the problem at hand. Finally, the super-Lagrangian refinement strategy we have used in this project allows in principle to reach smaller separations than possible with an SPH code in a realistic galaxy merger at a reasonable computational cost.


The paper is organized as follows. In section \ref{sec:simu_params}, we describe the hydrodynamical simulations we have performed. Results on SMBHs pairing and dynamical friction due to gas are then presented in section \ref{sec:results}. Finally, section \ref{sec:disc_concl} is devoted to discussion and conclusions.

\section{Hydrodynamical simulation parameters}
\label{sec:simu_params}

We use the AMR hydrodynamical code \RAMSES \citep{Teyssier2002} to model the evolution of a galactic major merger in which each galaxy hosts a SMBH at its center.
The galactic model, inspired by the one from \cite{MayerKazantzidis2008}, consists in a NFW isotropic dark matter halo, a rotating exponential stellar disc, a Hernquist bulge, a thin rotating exponential gas disc and a single motionless and non-accreting SMBH particle placed at the center of the galaxy.

The dark matter halo has a virial rotation velocity $v_{200} = 138 \kms$ ($M_{200}=8.7\times10^{11} \; \Msun$), a concentration parameter $c = 9.0$, a dimensionless spin parameter $\lambda=0.05$ and is made of $N_{dm}\sim7\times10^{6}$ particles.
The disk has a mass $M_{d}=0.05 \times M_{200}$, a scale length $R_{d} = 3.6 \; \textrm{kpc}$, a thickness $H_{d}=0.1 \times R_{d}$ and is made of $N_{*}=1.8\times10^{6}$ particles. The bulge has a mass $M_{b}=0.02 \times M_{200}$, a scale radius $a=0.1 \times R_{d}$ and is made of $N_{b}=8\times10^{5}$ particles.
As in  \cite{Kazantzidis2005}, the mass of the SMBH particle $M_{bh}=2.5\times10^{6} \; \Msun$ is chosen consistently with the $M_{bh}-\sigma$ relation \citep{FerrareseMerritt2000, Tremaine2002}. The gaseous disk is initialized on the AMR grid as a continuous density field with a total mass $M_{g}=4.18\times10^{9} \; \Msun$. 

For the merger model, we use the same parabolic orbit of a coplanar prograde galaxy encounter presented in \citet{MayerKazantzidis2008}. Initially, the two halos were seperated by twice their virial radii and their relative velocity was determined from the corresponding Keplerian orbit of two point masses. The distance at first pericenter is $b\simeq 15 \; \textrm{kpc}$ ($0.2\times \textrm{r}_{\mattext{vir}}$).
During the interaction, both SMBH particles stay quite nicely at the center of the galactic cores in which they were first embedded. After a few pericentric passages, dynamical friction finally makes the two galaxies merge at $t\simeq 5.244 \; \textrm{Gyr}$. The first $4 \; \textrm{Gyr}$ of the simulation have been run at low resolution, to let us spend most of the computational time during the final merger and follow the dynamics of the galactic cores at much higher resolution.

To avoid binary relaxation among DM/stellar particles, the gravitational softening of every particle except for the two SMBH particles cannot be smaller than $\epsilon_{min} = 3 \; \textrm{pc}$, consistently with our particle mass resolution. Meanwhile, the two SMBH particle dynamics and the gas dynamics follow the local resolution of the AMR grid without any limitation.

\subsection{Gas physics and equation-of-state}

A proper treatment of the thermodynamics of the multiphase interstellar medium is clearly beyond the scope of this paper.
This would require a much better resolution throughout the 2 colliding disks, as well as a complex cooling and heating model.
In this context, stellar feedback would play a crucial role, and our understanding of its effects on the structure of the turbulent ISM 
is still a matter of intense research, especially in the context of nuclear galactic disk. The seminal work of \cite{Wada2002} and \cite{Wada2009} have shown that such extreme environments give rise to a very wide range of gas temperature (from $10$~K to $10^8$~K) and densities, and velocity dispersions from 20~km/s up to 60~km/s, therefore significantly larger than in more quiescent disk conditions. This has led subsequent work on galaxy mergers to consider an effective equation of state for the gas, using a larger effective temperature than in normal galactic conditions, so that the properties of this extreme multiphase turbulent medium can be captured more realistically \citep{Robertson2005}. Very recently, this simplified approach was validated by \cite{Hopkins2012}, using a very high resolution simulation of a merging galaxy system with a sophisticated model for star formation and feedback.

In the same spirit as \cite{Springel2005} and \citep{Robertson2005}, we adopt  here also a  thermodynamical model based on a polytropic equation of state. As in \citet{Teyssier2010Antennae} and \citet{Bournaud2010TurbLMC},
we use for the hot virialized halo, defined as $n_H < n_{\rm h}$ with $n_{\rm h} = 10^{-3} \; \textrm{H/cc}$), a polytrope with $T=T_{\rm h} \left( n_H/n_{\rm h}\right)^{\Gammah - 1}$ with $T_{\rm h} = 4 \times 10^{6} \;\textrm{K}$ and $\Gammah=5/3$.
For the disk, defined as  $n_H > n_{\rm h}$, we use first  an isothermal equation-of-state with $T = 10^{4} \; \textrm{K}$.
At the end of the merger, the two black holes are embedded in the nuclear gas core of the remnant galaxy. The thermodynamics and the turbulent state of this inner region is the key parameter of such a study. The most important property we would like to capture is the large turbulent velocity providing pressure support in the inner region. Using as characteristic density $n_{\rm d} =1000\; \textrm{H/cc}$, we use the following equation-of-state across the whole disk 
\begin{equation}
T={\rm max} \left[ 10^4~{\rm K}, T_{\rm d} \left( n_H/n_{\rm d}\right)^{\Gammad - 1}\right]
\end{equation}
As shown by \citet{Escala2005} and \citet{Dotti2007}, an {\it effective} sound speed in the range $60-80 \kms$ corresponds to a pressure scale height comparable to that deduced from observations of circumnuclear disks in nearby galaxies \citep{DownesSolomon1998}. In our model, we adopted a effective temperature $T_{\rm d} = 3 \times 10^5 \; \textrm{K}$ or equivalently a effective sound speed $c_{\mattext{s},{\rm d}} = 75 \kms$. 

It is interesting to investigate the impact of various models on the black hole orbital decay. For that purpose, we consider first a model which corresponds to turbulent gas with constant effective entropy, so that $\Gammad=5/3$. On the other hand, 
\cite{SpaansSilk2000} and \cite{Klessen2007} showed that the thermodynamic state of a solar metallicity gas heated by a starburst can be well approximated by an ideal gas with adiabatic index $\Gammad = 1.3 - 1.4$ over a wide range of densities.
Consequently, we also adopted a second model for which the turbulent high-density gas is more dissipative than in the first model and follow a $\Gammad=7/5$ polytropic equation-of-state.



\section{Results}
\label{sec:results}

\begin{figure}
	\centering
	\includegraphics[width=0.23\textwidth]{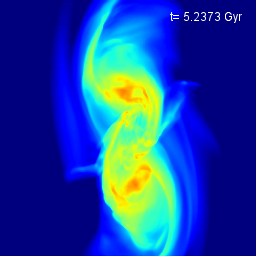}
	\includegraphics[width=0.23\textwidth]{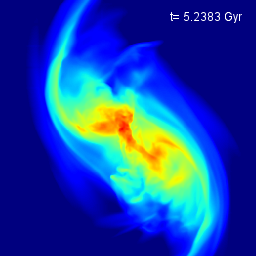}\\
	\includegraphics[width=0.23\textwidth]{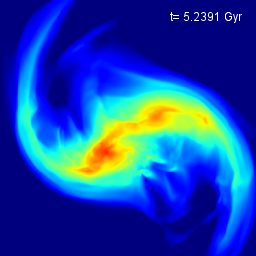}
	\includegraphics[width=0.23\textwidth]{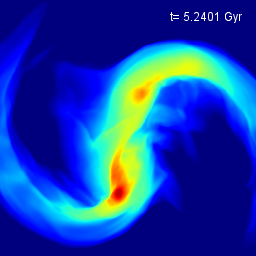}\\
	\includegraphics[width=0.23\textwidth]{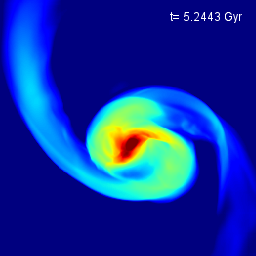}
	\includegraphics[width=0.23\textwidth]{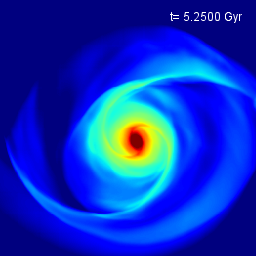}
	\caption{Mass-weighted gas density maps during last pericenter and final merger. The line-of-sight is perpendicular to the orbital plane and the maps are $1.8 \; \textrm{kpc}$ wide. While fairly symmetric before the pericenter (top left), the density distribution becomes clearly asymmetric after the pericenter (middle). After the final merger, a thick $\sim 10^{9}\;\Msun$ gaseous nuclear disk is formed (bottom), in which the two SMBHs start orbiting.}
	\label{fig:asym_shock}
\end{figure}

A few $\Myr$ before the final merger of the two gaseous cores, during the last pericenter, the symmetry of the gas distribution changes dramatically.
In Figure~\ref{fig:asym_shock}, one can see that the gas distribution, perfectly symmetric before the last pericenter, becomes asymmetric afterwards.
This is due to small scale turbulence between the two galactic cores (top left view) that made the shocks within the gas asymmetric during the last pericenter.
Afterwards, one of the galactic core (and SMBH within) ends up with a bit slower and with a little more mass than the other (bottom right view).
An asymmetrical SMBH injection in the nuclear disk results from this symmetry breaking : one of the black hole is injected a few parsecs away from the nuclear disk center, while the other starts orbiting with an apocenter as large as $r\sim 100 \; \textrm{pc}$.

\subsection{Nuclear disk formation and turbulence dissipation}

\begin{figure*}
	\centering
	\includegraphics[width=0.33\textwidth]{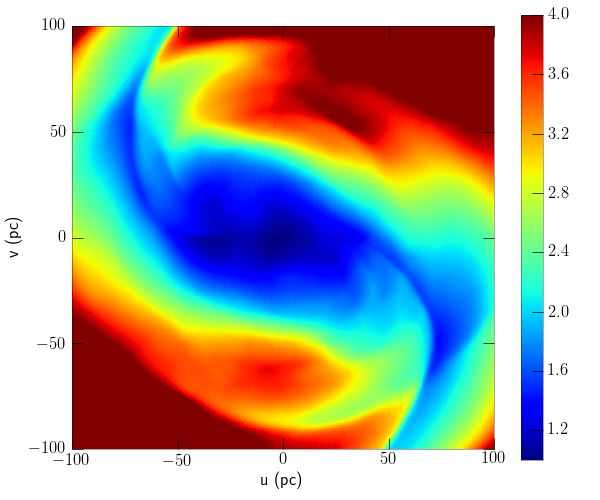}
	\includegraphics[width=0.33\textwidth]{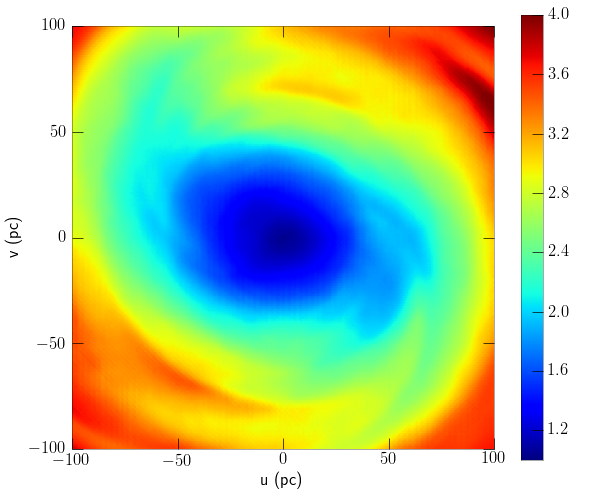}
	\includegraphics[width=0.32\textwidth]{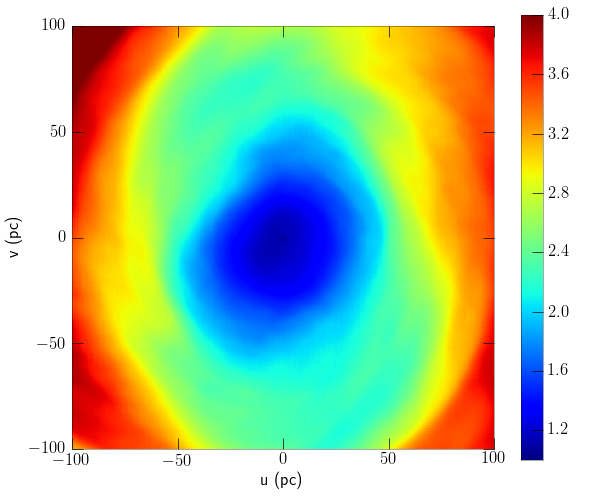}
	\includegraphics[width=0.33\textwidth]{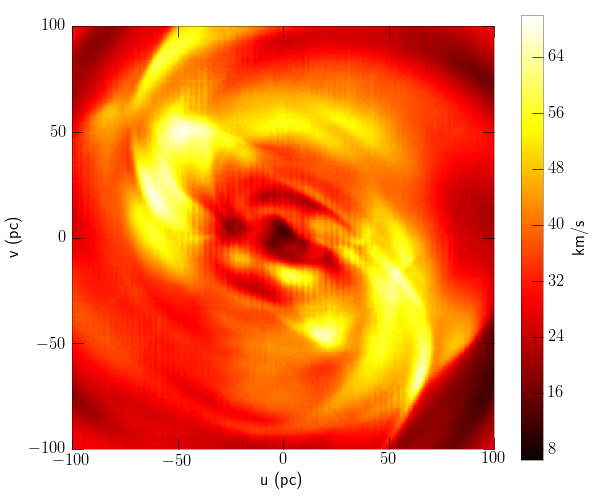}
	\includegraphics[width=0.33\textwidth]{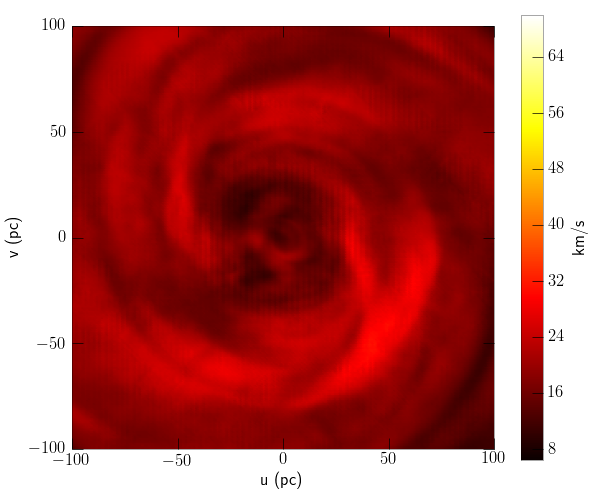}
	\includegraphics[width=0.32\textwidth]{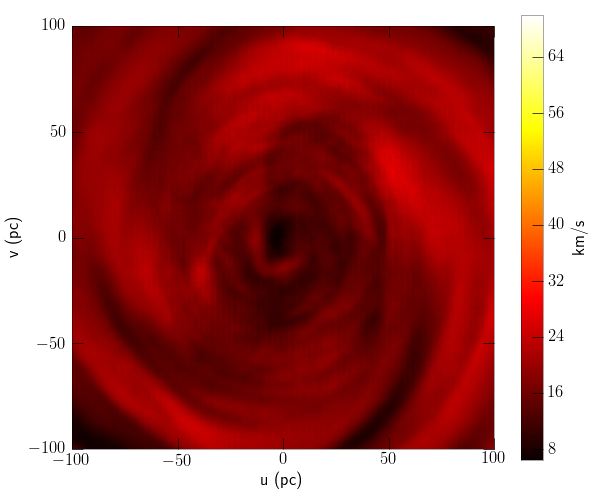}\\
	\includegraphics[width=0.33\textwidth]{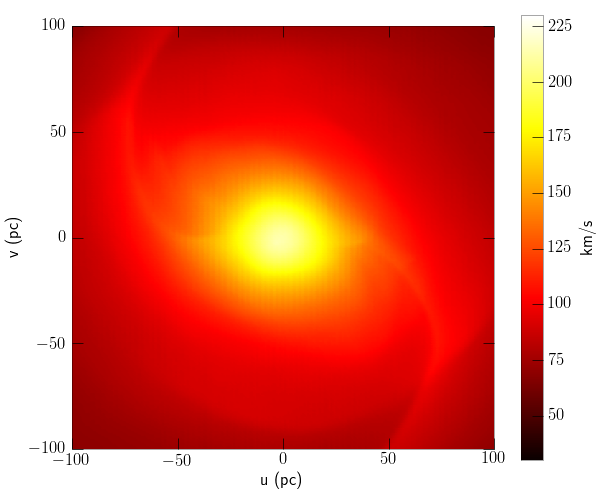}
	\includegraphics[width=0.33\textwidth]{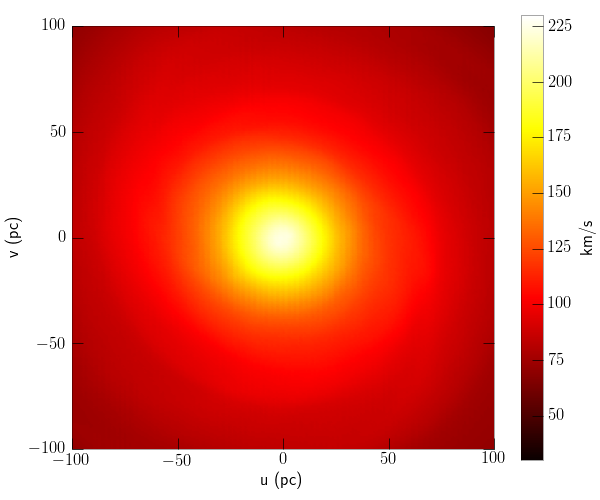}
	\includegraphics[width=0.32\textwidth]{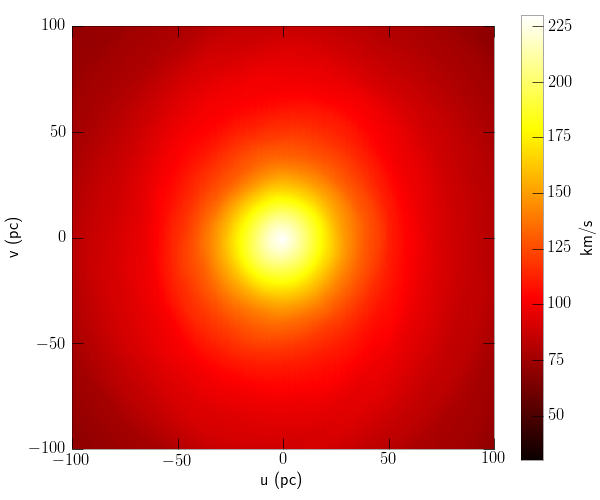}\\
	\includegraphics[width=0.33\textwidth]{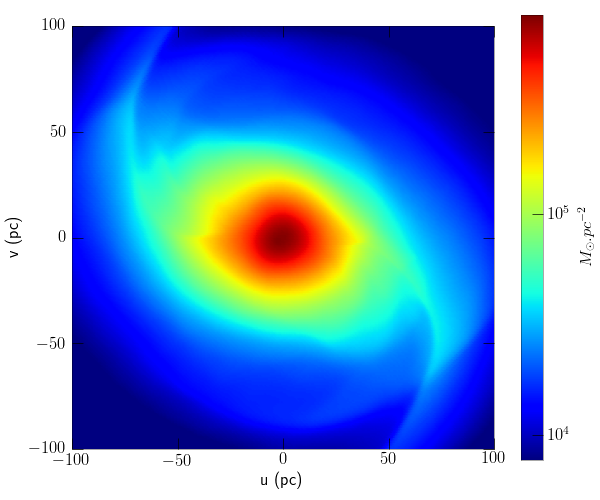}
	\includegraphics[width=0.33\textwidth]{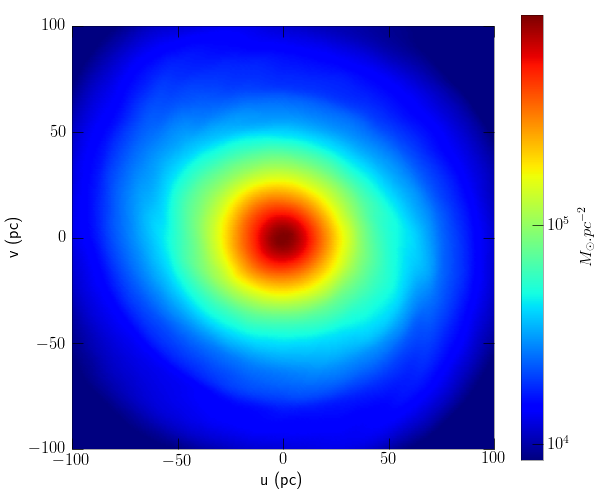}
	\includegraphics[width=0.32\textwidth]{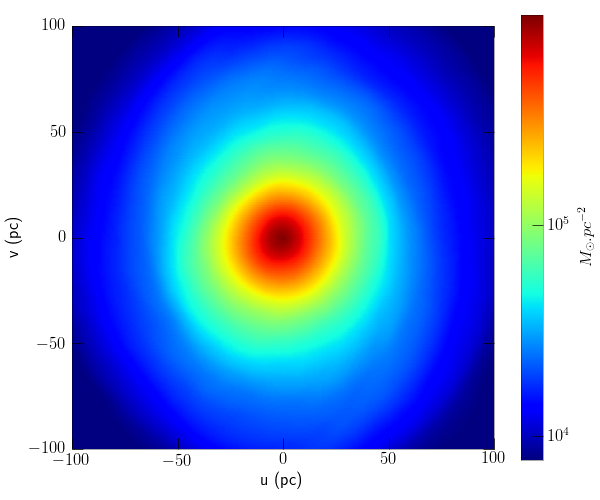}\\
	\includegraphics[width=0.33\textwidth]{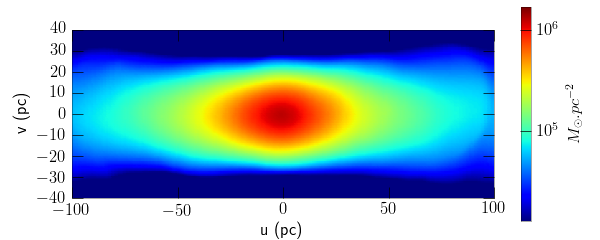}
	\includegraphics[width=0.33\textwidth]{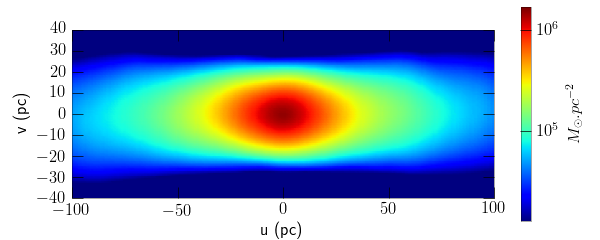}
	\includegraphics[width=0.32\textwidth]{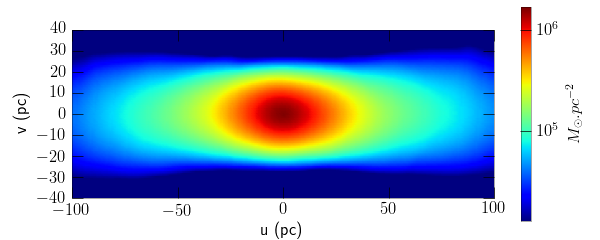}\\
	\caption{200 \textrm{pc}-wide nuclear disk ($\Gammad=7/5$ case) at $t (\textrm{Gyr})=5.25$ (left), $5.256$ (middle) $5.2571$ (right). From top to bottom are shown the Toomre parameter, velocity dispersion, sound speed, face-on and edge-on gas surface density maps.}
	\label{fig:nuc_disk}
\end{figure*}

 
After an additional $4 \Myr$ , the two galactic cores finally merge together. In the $\Gammad=5/3$ case, the core coalescence leads to the formation of a $\sim 140 \; \textrm{pc}$ thick gas spheroid, while in the $\Gammad=7/5$ case, the nuclear region is more disky with a disk thickness of $\sim 60 \; \textrm{pc}$. In both cases, the enclosed gas mass within $100 \; \textrm{pc}$ is $ \sim 10^{9} \; \Msun$.
The parameters of the nuclear disk we obtain from this galaxy merger simulation are consistent with nuclear regions observed in Ultra Luminous Infrared Galaxies \citep[ULIRGs;][]{DownesSolomon1998, SandersMirabel1996}, except that our model is missing the intense star formation processes of the observed nuclear regions.
Figure~\ref{fig:nuc_disk} shows the velocity dispersion, sound speed, surface density and Toomre stability parameter maps of this disk ($\Gammad=7/5$ case). The Toomre parameter is above $1$ across the whole disk. The velocity dispersion $\sigma_{\mattext{v}}$ is significantly smaller than the sound speed $\csound$.

Compared to SPH calculations made by \citet{MayerKazantzidis2008} with the same merger model, the overall properties of the disk are qualitatively in good agreement with our AMR simulations.
The nuclear disk formed in the SPH simulation was however found to be marginally gravitationally unstable. This caused the formation of a strong spiral wave and the collapse of the disk as the angular momentum was driven outward. The resulting inflow was at the origin of a fast decrease of the ambient gas density around the SMBH and of the stalling of the orbital decay. 
Testing the robustness of this result using a different thermodynamical model and a different numerical method is precisely the main justification of the present work. A more stable nuclear disk configuration would result in a larger gas density and in a stronger drag force.
In our AMR model, without any other source of turbulence than gravity, the gas velocity dispersion is quite high at the beginning, and a rather strong spiral mode can be seen in the density maps.
But both the spiral wave and the turbulence are slowly dissipated from $\sigma_v \simeq 70 \kms$ to $\sigma_v \simeq 30 \kms$ (Fig.~\ref{fig:nuc_disk}, second row) in the nuclear disk, over a timescale $t_{\mattext{cross}}=h/\sigma_v \simeq 2 \Myr$.
In contrast to the SPH results, we do not see a sustained transport of angular momentum outward, and the disk settles in a stable pressure equilibrium, with high gas density but also high gas sound speed ($\csound \sim 200 \kms$).


\subsection{SMBHs orbital decay}

\begin{figure}
	\centering
	\includegraphics[width=0.5\textwidth]{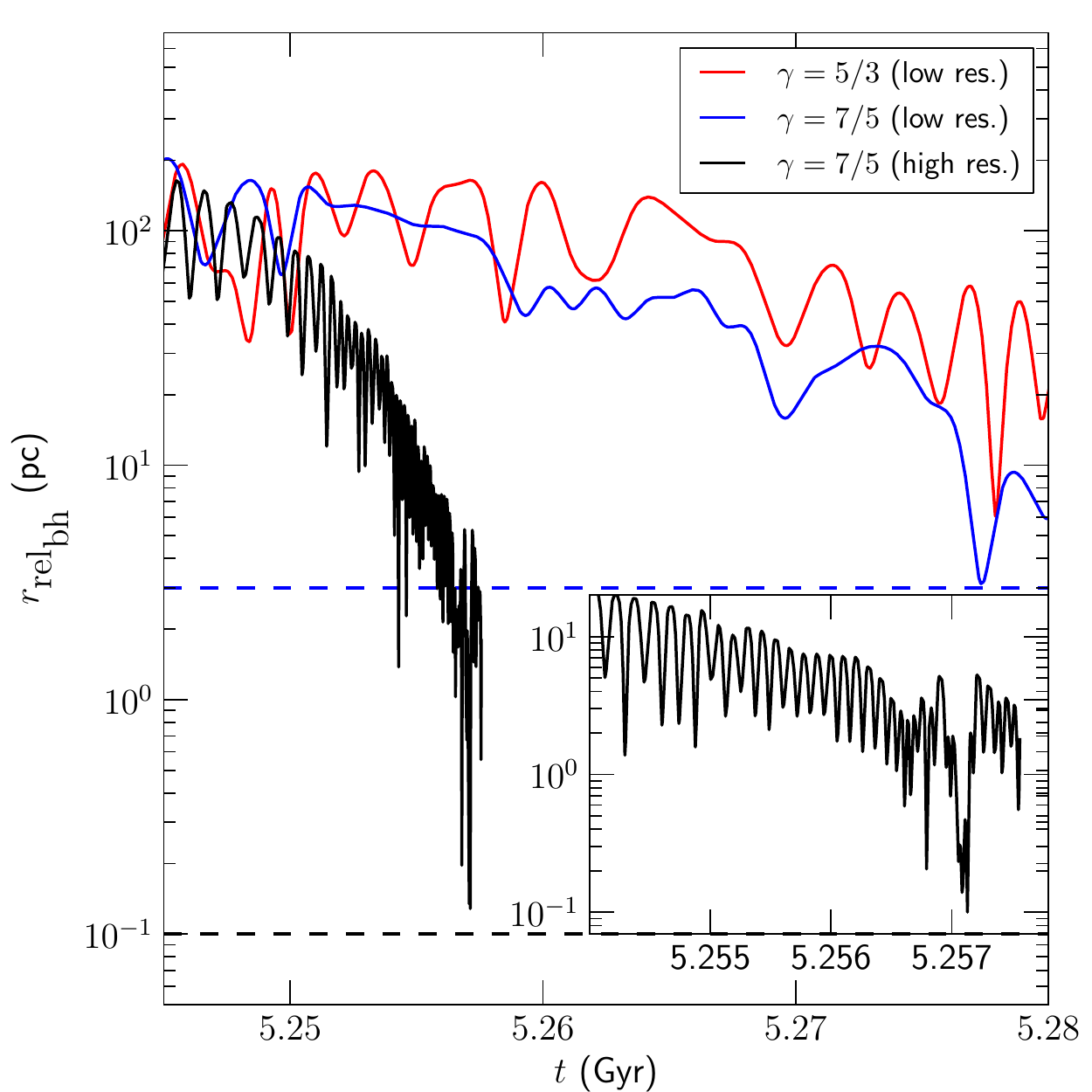}
	\caption{SMBH particles relative separation evolution for the low-resolution runs ($\Delta x_{min}=3$ pc) with $\Gammad=5/3$ (red) and $\Gammad=7/5$ (blue) and the high-resolution run ($\Delta x_{min}=0.1$ pc) with $\Gammad=7/5$ (black). The blue (resp. black) dashed line corresponds to the low (resp. high) spatial resolution limit. The red dashed line corresponds to the black hole separation limit where the binary becomes hard ($M_{\mattext{binary}}=2 M_{\mattext{BH}}=M(r<3 \; \textrm{pc})$).}
	\label{fig:rrel_bh}
\end{figure}

We follow the evolution of the two SMBH particles in this nuclear region for the different models. 
Figure~\ref{fig:rrel_bh} shows the evolution of the black hole relative separation. The low resolution runs are in good agreement with \cite{MayerKazantzidis2008} results. While the hot ($\Gammad=5/3$) thermodynamical model leads to a stalling of the black hole orbital decay around $r \sim 40 \; \textrm{pc}$, the cold model ($\Gammad=7/5$) let the black hole relative separation falls down to the numerical resolution ($\Delta x=3 \; \textrm{pc}$) after only 30 $\Myr$.
These first experiments are however not conclusive, since, as we will demonstrate in the next section, this resolution is too low to resolve properly the wake causing the hydrodynamical friction in the nuclear disk.

In our high resolution run, shown as the black line in Figure~\ref{fig:rrel_bh}, the environment of the SMBH is much better resolved. 
As a consequence, the SMBH relative separation decrease from $100 \; \textrm{pc}$ down to $1 \; \textrm{pc}$ in less than $10 \Myr$. 
Although the black hole binary system reach very briefly a relative separation close to our resolution limit ($\Delta x =0.1 \; \textrm{pc}$), the orbital separation of the black holes is stalling well above the resolution limit and settles around 2 parsecs where the binary system becomes hard ($2 M_{\mattext{BH}}=M(r<3 \; \textrm{pc})$, red dashed line).
Using a different thermodynamical model than the previous SPH simulations results in a denser, more stable nuclear disk, but, as we now show in more details, this also leads to inefficient hydrodynamical friction and failure of the model to harden the binary system down to sub-parsec scales in the center of this nuclear disk.
Note however that, in our case, the dynamical friction time scale is increased by roughly one order of magnitude, so that orbital decay is not stopped entirely and SMBH pairing will probably take place after several tens of  $\Myr$. We couldn't follow the late time evolution 
over such a long period by lack of sufficient computational resources.  

\subsection{Dynamical friction in a gaseous medium}

\begin{figure*}
	\includegraphics[width=\textwidth]{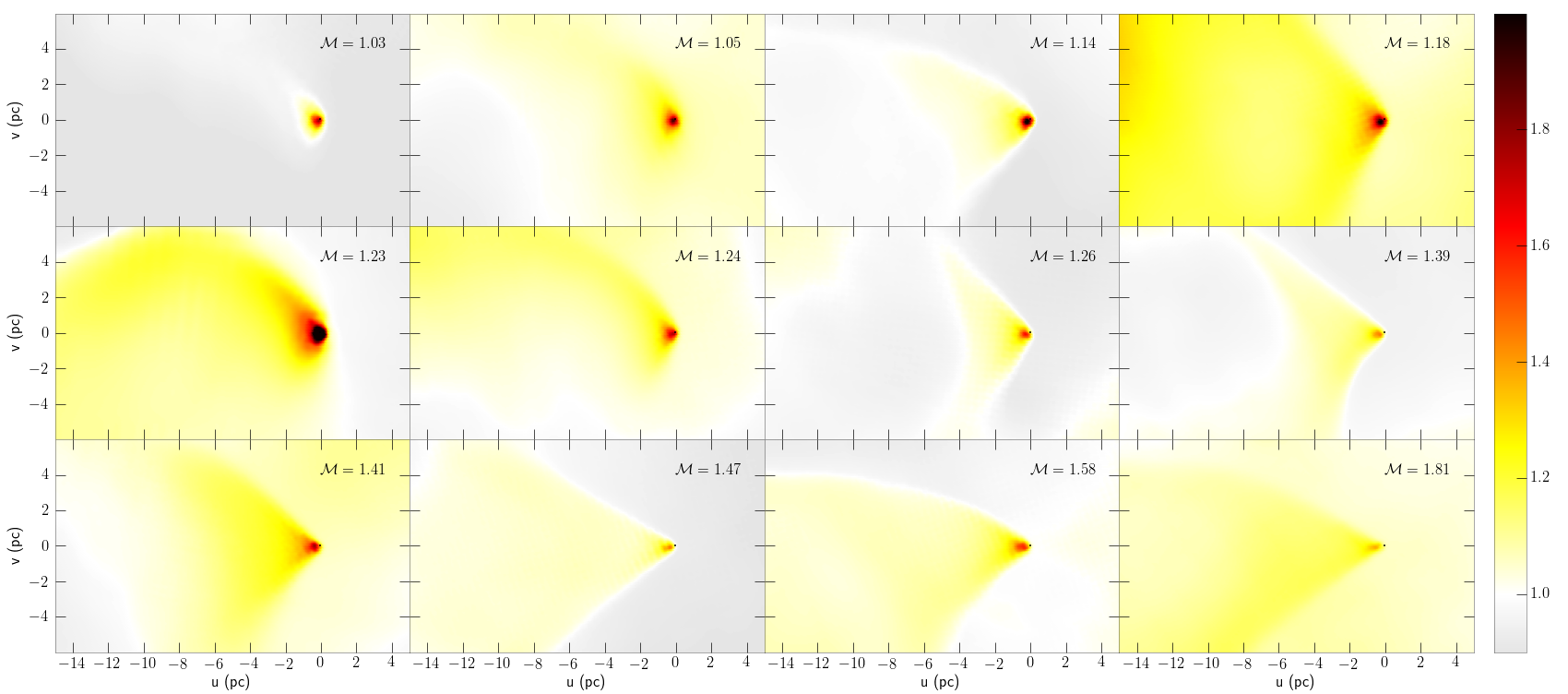}
	\caption{Gas Overdensity maps behind the SMBH particle (the relative velocity of the SMBH compared to the surrounding gas is pointing rightward). The hydrodynamical wakes and Mach cones are shown for increasing values of $\Mach=V_{\mattext{bh/gas}} / \csound$. These wakes are clearly stronger and make the dynamical friction much more efficient when the black hole is in a transonic regime ($\Mach=1.14, 1.18, 1.23$).}
	\label{fig:hydro_wake}
\end{figure*}

\begin{figure}
	\centering
	\includegraphics[width=0.5\textwidth]{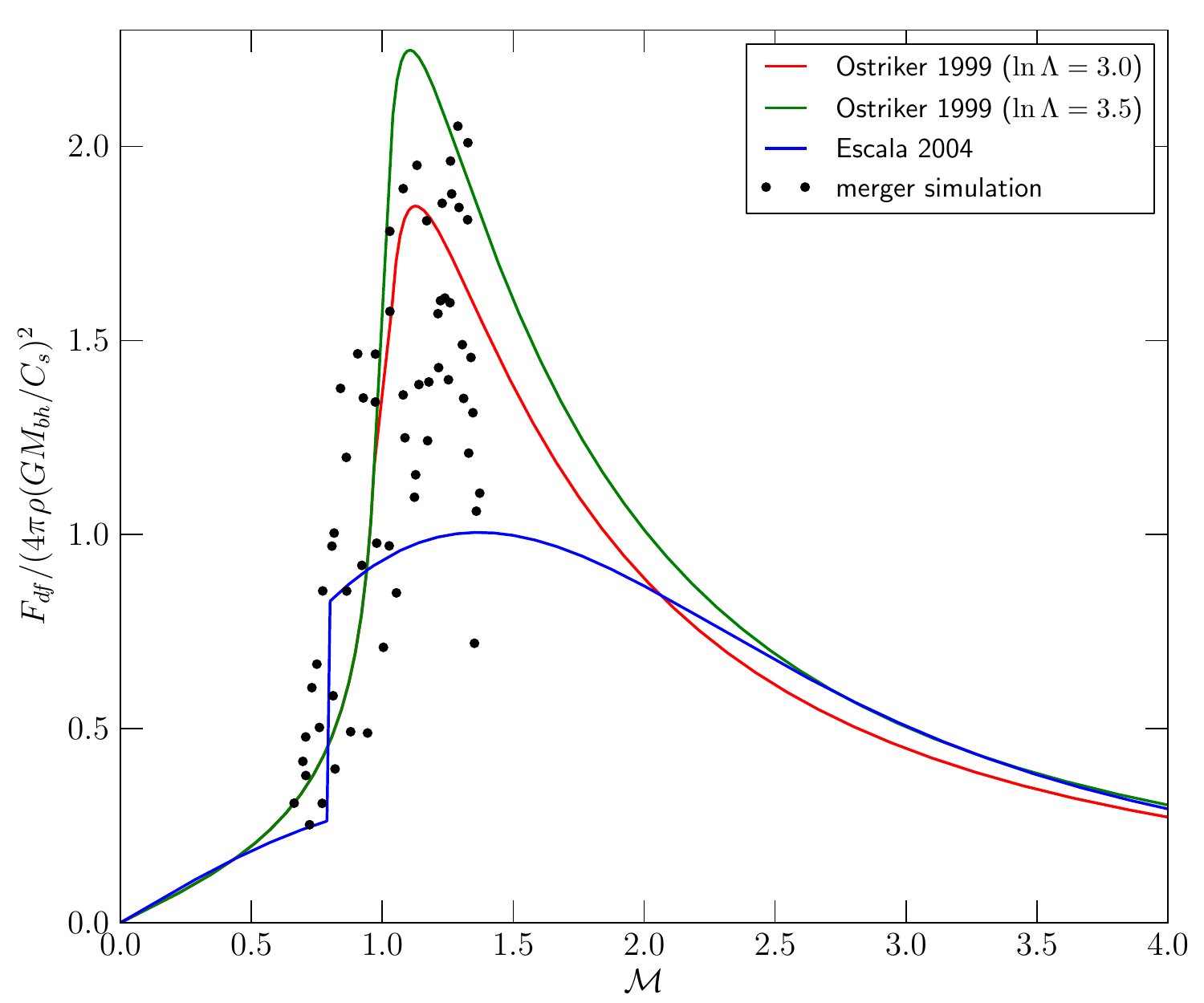}
	\caption{Dimensionless factor $\fgas$ of the dynamical friction force as a function of the Mach number $\Mach=v_{bh/gas}/\csound$. Analytical prediction by \citet{Ostriker1999} for a Coulomb logarithm $\ln \Lambda = 3$ (red) and $\ln \Lambda = 3.5$ (green) compared to numerical results from \citet{Escala2004} (blue) and this work (black dots), corresponding to a numerical Coulomb logarithm $\ln \Lambda \simeq 3.2$. }
	\label{fig:Mach_fgas}
\end{figure}

Figure~\ref{fig:hydro_wake} shows the gas overdensity induced by a SMBH particle during an orbit where the black hole is about $r = 20 \; \textrm{pc}$ from the center of the nuclear disk. On each panel, the position of the black hole is marked by a black dot and the orientation is taken so that the relative velocity of the black hole in its surrounding gas is pointing rightward.
The black hole induces a shock and a trailing hydrodynamical wake, which both get stronger as the black hole reach a transonic regime where the Mach number is defined using the relative velocity of the SMBH with respect to the gas
\begin{equation}
	\Mach = \frac{v_{\mattext{bh}}-v_{\mattext{gas}}(\mathbf{x}_{\mattext{bh}})}{\csound(\mathbf{x}_{\mattext{bh}})}
\end{equation}
The hydrodynamical wake exerts a gravitational drag on the black hole proportional to $4\pi \rho \csound^{2} R_{\rm BH}^{2}$, where the Bondi radius is expressed as $R_{\rm BH} = GM_{\rm BH}/\csound^{2}$ \citep{Ruffert1996}. The true efficiency of the dynamical friction exerted by the gaseous medium on the black hole can be expressed by the dimensionless correction factor 
\begin{equation}
	\label{eq:fgas}
	\fgas=\frac{F_{\mattext{DF}} }{4\pi\rho(GM_{\mattext{BH}}/ \csound)^{2}}
\end{equation}
The analytical study by \citet{Ostriker1999} provided a formula for $\fgas$ 
\begin{eqnarray}
	\label{eq:fgas_Ostriker1999}
	\fgas_{\mattext{subsonic}} &=& \frac{1}{\Mach^{2}} \left[ \frac{1}{2} \ln \left( \frac{1+\Mach}{1-\Mach} \right) - \Mach\right]\\
	\fgas_{\mattext{supersonic}} &=& \frac{1}{\Mach^{2}} \left[  \frac{1}{2} \ln \left( \Mach^{2}-1 \right) +  \ln \Lambda \right]
\end{eqnarray}
where the Coulomb logarithm $\ln \Lambda = \ln(r_{\mattext{max}} / r_{\mattext{min}})$ accounts for the maximum and minimum radial contributions to the drag, as in the standard case of dynamical friction exerted by a stellar background \citep{Chandrasekhar1943}.
Note that previous numerical studies have claimed that these analytical formulae were probably overestimating the effect of dynamical friction for the transonic regime \citep{Escala2004, Sanchez2001}.
In order to compare our numerical drag to these analytical and numerical estimates, we computed the gravitational force of the perturbed density field seen by the orbiting black hole particle.
Figure~\ref{fig:Mach_fgas} shows the dynamical friction dimensionless factor $\fgas$. The numerical values obtained during one orbit of the distant black hole particle is in good agreement with the analytical prediction of \cite{Ostriker1999} if we use $\ln \Lambda \simeq 3$.
Deviations from the analytical model could be easily explained by the time-dependent nature of the SMBH orbit, while the  theory of \citet{Ostriker1999} is based on a strictly stationary flow around the black hole.

\begin{figure}
	\centering
	\includegraphics[width=0.5\textwidth]{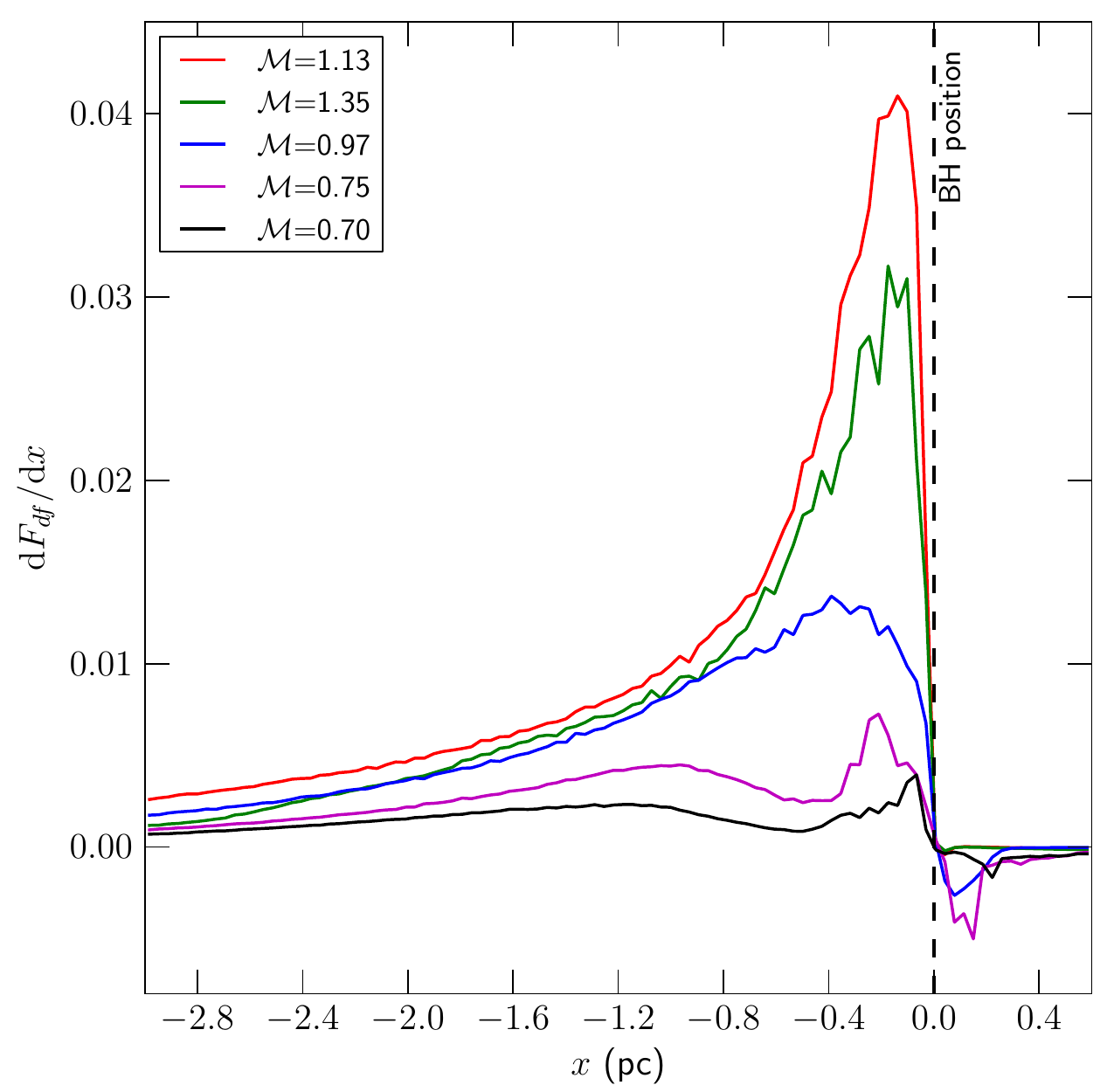}
	\caption{Spatial contribution to the gravitational drag the gas exerts on the black hole particle at different Mach number. The strength of the drag peaks around transonic regime ($\Mach=1.13$, red).Whatever the Mach number, most of the gravitational drag comes from the $2-3 \; \textrm{pc}$ region trailing behind the black hole.}
	\label{fig:df_location_profile}
\end{figure}

In Figure~\ref{fig:df_location_profile}, we plot the fractional contribution to the total gravitational drag of plan-parallel slabs of gas perpendicular to the propagation axis of the BH. In this plot, the x-coordinate is the distance of each slab (of size $3 \;\textrm{pc}$) to the BH. 
We see immediately that for various Mach numbers, the maximum radius that contributes to the drag force is roughly $r_{\rm max} \simeq 2.5$ pc.
Moreover, the fractional drag profile is well resolved by the cell size of our simulation, which quite naturally corresponds to the minimum scale that contributes to the drag $r_{\rm min} \simeq 0.1$ pc. From these two numbers, we can estimate the Coulomb logarithm in our simulation as $\ln \Lambda = \ln r_{\rm max} / r_{\rm min} \simeq \ln 25 \simeq 3.2$.
We see in Figure~\ref{fig:Mach_fgas} that the analytical model of \citet{Ostriker1999} using $\ln \Lambda=3$ (red) or $\ln \Lambda=3.5$ (green) is a perfectly good fit to our numerical data. From Figure~\ref{fig:df_location_profile}, we see that resolving the SMBH environment with sub-parsec resolution is mandatory in order to resolve the wake properly. At Mach number slightly above unity, the wake profile sharply declines towards the BH position, and even higher resolutions would probably result in a stronger drag force, by effectively raising the value of the Coulomb logarithm.
 
\subsection{A transition from fast to slow orbital decay}

Let us consider a simple model of orbital decay. Assuming that the black hole evolves on circular orbits with circular velocity $v_{\mattext{circ}}(r)$ 
and undergoes a hydrodynamical friction force opposite to its velocity vector and of amplitude $F_{\mattext{df}}$.
The BH angular momentum writes
\begin{equation}
L= M_{\mattext{BH}} r v_{\mattext{circ}} 
\end{equation}
and the angular momentum losses write
\begin{equation}
	\dot{L} = | \mathbf{r} \times \mathbf{F_{\mattext{df}} }| = - r F_{\mattext{df}}
\end{equation}
We can derive the orbital decay timescale as 
\begin{equation}
	\label{eq:decay_timescale}
	\tauOD = \frac{L}{\dot{L}} = \frac{M_{\mattext{BH}} v_{\mattext{circ}} }{F_{\mattext{df}} }
\end{equation}
When the BH is evolving at large radii, say $20 \; \textrm{pc}$, the gas density is rather low, $\rho=6\times10^{5}  \Hcc$, as is the gas sound speed with $c_s=230 \kms$. 
The Mach number of the BH relative speed is however measured in the simulation to be Mach $\simeq 1.2$, which, using the \citet{Ostriker1999} formula into Equation~\ref{eq:decay_timescale} results in a rather fast orbital decay with $\tauOD(r=20 \; \textrm{pc})\simeq 1 \; \textrm{Myr}$, quite consistent with the orbital evolution of the BH seen in Figure~\ref{fig:rrel_bh}.
When the black hole fall within the innermost $r=5\; \textrm{pc}$ of the nuclear disk, its relative velocity drop below $100 \kms$ while the local sound speed reach $\csound=310 \kms$. In this strongly subsonic regime, the drop of $\fgas(r<5 \; \textrm{pc})\simeq 0.05$ is not compensated by the increase of the ratio $\rho/\csound^{2}$ and the orbital decay timescale rises to $\tauOD(r\sim 1 \; \textrm{pc})\simeq 10 \; \textrm{Myr}$.
This, together with the fact that the gas mass present between the two black holes is comparable or smaller than the sum of the masses of the two holes below pc scale separations, explains in principle why the decay becomes inefficient. We will return on this point in the Discussion below.
While the BH is sinking towards the center of the nuclear disk, its orbit is getting more and more circular, as found by \citet{Dotti2006, Dotti2007}.
This will reduce the relative velocity of the BH with respect to the gas disk even more, and strengthens the robustness of our conclusion, namely that BH pairing in a stable, pressure-supported, nuclear disk is a rather slow mechanism.

The issue of gap-opening is an important point in the process of black hole binary formation and its final coalescence. If the black holes are  able to open a circumbinary gap, the density of the gas surrounding the black holes drops significantly and dynamical friction is no longer efficient to make the binary separation decrease even further.
Decay might continue if there are other viscous processes acting in the disk, as in the case of migration in protoplanetary disks, but the timescale will be related to the specific nature of the viscous process (e.g. transport of angular momentum via MHD or density waves supported by the self-gravity of the gas) which we either do not include (e.g. MHD instabilities) or cannot model properly at scales near our resolution limit (e.g. density waves). 
One can determine the gap-opening criterion by comparing the gap-closing and gap-opening times as it has been done in the context of planetary rings \citep{GoldreichTremaine1982}.
In our nuclear disk of thickness $h=60 \; \textrm{pc}$, the black hole mass gap-opening criterion can be written \citep[][eq.~6]{Escala2005}
\begin{equation}
	M_{\mattext{BH}} \ge \sqrt{\frac{\Delta r / h}{4 \pi \Mach \fgas(\Mach)}} \left( \frac{h}{r} \right) M(<r)
\end{equation}
When the two black holes fall into the innermost $5 \; \textrm{pc}$, $\Mach\simeq0.15$ and $f^{\mattext{(gas)}}(\Mach) \simeq 0.05$. As shown in the previous section, the black hole should clear a gap of typical size $\Delta r \simeq 2.5 \; \textrm{pc}$ to prevent the formation of a hydrodynamical wake and make the dynamical friction inefficient. Consequently, the black hole critical mass is $2\times10^{8} \; \Msun$, well above the mass of our black hole particle which could never open a gap in such a nuclear disk. Even if a Bondi-Hoyle accretion modeled was taken into account in the simulation, \citet{Callegari2011} showed that, in minor merger simulations, one of the black hole could grow by nearly an order of magnitude in mass at the most, which would not be enough for our $2.5\times10^{6} \; \Msun$ black hole to reach the critical mass.
However, since the observed black hole masses range from $10^{6}$ to a few $10^{9} \; \Msun$, the formation of a circumbinary gap by a more massive black hole is possible in our disk.



\section{Conclusions and discussion}
\label{sec:disc_concl}

\subsection{Comparison with previous Smoothed Particle Hydrodynamics simulations}

Our numerical study, based on the Adaptive Mesh Refinement technique, addresses  the formation of a nuclear gas disk resulting from an equal-mass dissipative galaxy merger and the concurrent formation of a SMBH binary at parsec scales within such a disk.
Our results confirm previous findings of SPH simulations with the \GASOLINE code, with particle splitting that started from very similar initial conditions \citep{MayerSci2007, MayerKazantzidis2008}.
After the initial asymmetric black hole injection following the final merger of the galactic cores, the dynamical friction from the gaseous background dominates the one due to the stellar background and makes the black hole separation falls from $100 \; \textrm{pc}$ to a few parsecs in less than $10 \Myr$, in qualitatively good agreement with \citet{MayerSci2007} results.
Indeed in the reference simulation adopting a polytropic equation of state with $\Gammad = 7/5$, the mass, size and typical density of the nuclear disk that arises (respectively, $\sim 10^9 M_{\odot}$ and $60 \;\textrm{pc}$) are extremely close to those of the nuclear disk in \citet{MayerSci2007}, as is its characteristic temperature. These properties are also consistent with observed nuclear regions in ULIRGs. 
Being embedded in a background with density and temperature similar to that in the \citet{MayerSci2007} calculations, the pair of SMBHs sinks by dynamical friction very rapidly after the merger, and becomes bound also at a similar separation, about $6 \;\textrm{pc}$. 
We have also run two lower resolution simulations with different polytropic indexes, one with $\Gammad=7/5$ as in the reference run and another one with $\Gammad = 5/3$; the comparison shows the same trend found by \citet{MayerSci2007}, namely that with a stiffer EOS ($\Gammad=5/3$) the orbital decay is slower, both because the decay regime is more subsonic and because the gaseous background has a lower characteristic density, both effects going in the direction of reducing dynamical friction.

At a closer inspection, there are however some differences between the SPH nuclear disks of \citet{MayerSci2007} and that in the new \RAMSES simulations presented in this paper. First, the disk in the SPH simulations displayed a stronger spiral structure since its appearance, as a result of a higher central surface density and thus a stronger self-gravity (the Toomre parameter was close to 1.5 while $\ge 2.0$ in our disk, see top row of fig. \ref{fig:nuc_disk}).
The stronger spiral modes lead to a more effective transfer of angular momentum and thus a continuously increasing central density. Second, the level of non-thermal motions in the gas, usually termed "turbulence", was a factor of about 3 higher than in the \RAMSES run ($\sim 100 \kms$ instead of $30-40 \kms$), likely as a result of the stronger self-gravitating response.
The reason for these differences are not clear. The higher numerical viscosity in the SPH runs, due to the use of explicit Monaghan artificial viscosity, especially the quadratic term in shock dissipation during the merger, is expected to dissipate gas motions into thermal energy.
This would damp the turbulence faster, at odds with the higher turbulence seen in the SPH simulation, but would also generate viscous transport of angular momentum, leading to an increase in central density and therefore a higher susceptibility to gravitoturbulence, at least in the innermost region.
Another possible cause of the difference is the fact that in the \GASOLINE runs of  \citet{MayerSci2007} the energy equation was solved, including shock dissipation via artificial viscosity, possibly producing a nuclear disk with lower entropy immediately after the merger relative to the \RAMSES runs adopting a fixed polytropic equation of state which does not account, by construction, for entropy dissipation in shocks.
Finally, as we explain below, resolution in SPH and AMR runs is not guaranteed to be identical even if the setup is designed to be as close as possible as in the case that we are discussing.

\subsection{Resolving the drag force and the associated wake}

The exact timescale of the sinking of the SMBH binary is however dependent on resolution, as was also noticed in \citet{MayerKazantzidis2008}, in particular becomes shorter for increasing resolution.
At higher resolution the dynamical friction wake is better resolved as density gradients are better captured, this being a likely reason behind the faster dynamical friction timescale \citep{1986ApJ...300...93W}. In collisionless systems, an analogous resolution dependence of dynamical friction, mediated by gravitational softening, has been reported extensively in the literature \citep{1983ApJ...274L...1W,1989MNRAS.239..549W,ColpiMayerGovernato1999}.
The formation of a front shock and a trailing hydrodynamical wake which exerts a gravitational drag on the black hole, the so called dynamical friction wake, is a remarkable result of our \RAMSES simulations. The \RAMSES simulations presented in this paper are indeed the first that allow to capture the wake of dynamical friction very clearly in a highly dynamic situation such as that of a nuclear disk arising from merging galaxies. We show that this gravitational drag is due to overdense gas within $2-3 \; \textrm{pc}$ behind the black hole and that the efficiency of the hydrodynamical friction peaks at transonic regime, in fair agreement with the analytical prediction from \citet{Ostriker1999}, with a Coulomb logarithm close to our resolution limit, suggesting that the dynamical friction wake structure is not fully converged.

Our \RAMSES simulations, owing to the aggressive refinement enabled by the AMR technique, allow to reach a spatial resolution of $0.1 \; \textrm{pc}$ in the center  of the disk where  the SMBHs are sinking, which is ten times better than the nominal resolution in \citet{MayerSci2007}. We caution, however, that comparing the resolution in AMR and SPH is not straightforward.
In particular, in the SPH simulations of \citet{MayerSci2007} and \citet{MayerKazantzidis2008}, we adopted a fixed gravitational softening in the high resolution region after particle splitting, which can thus be considered the actual limit of spatial resolution in those calculations (the SPH smoothing length being smaller in high density regions due to the large number of particles employed).
In \RAMSES the gravitational force resolution is not fixed, rather it is tied to the cell size, therefore it shrinks as the refinement is applied to the cells , a situation more reminiscent of what happens with SPH codes adopting an adaptive softening length \citep[e.g.][]{BateBurkert1997}. 
We note that, in \citet{MayerKazantzidis2008}, a run was presented with a maximum spatial resolution (in terms of gravitational softening) comparable to the highest resolution simulation presented here. Although the simulation was carried out only for a few orbits after the black holes have formed a binary, rather than for many orbits as in the \RAMSES simulations presented here, in both cases the separation of the SMBHs appears to fluctuate significantly, with no clear signs of sustained decay below parsec separation.

\subsection{Is there a last parsec problem~?}

Below such separation it is expected that dynamical friction will become inefficient because the mass of the gaseous background enclosed within the orbit of the two SMBHs becomes smaller than the mass of the SMBH binary.
In addition, in the innermost $5 \; \textrm{pc}$ of the nuclear disk, the sound speed reaches $\sim 300 \kms$, a subsonic regime for the motion of SMBHs, which also implies an inefficient dynamical friction.
In polytropic equilibrium models of nuclear disks, \citet{Escala2004} reported an asymmetric torque due to an ellipsoidal deformation of the density distribution around the binary, which extracted angular momentum from the binary allowing it to shrink further.
This is not seen here nor in the previous SPH simulations of some of us. One reason might be that the thermodynamical conditions and density profile of the disk that develops here are different from those in equilibrium disk models.
In particular, the system analyzed by \citet{Escala2004} had a sound speed of about $60 \kms$ even at the center, hence it was a much colder gas disk than ours.
On the other hand, a colder disk could act against the decay in two ways, namely by driving a stronger mass inflow by gravitational torques that steepens the central density on a dynamical timescale and might thus evacuate the gas around the black hole binary, or might provide more favourable conditions for the opening of a gap by the binary after it has become bound.
Finally, it is important to note that binary decay stalls around $2-3 \; \textrm{pc}$, a separation at which the orbital evolution might not be correctly captured with only $\sim20$ AMR cells. A similar resolution problem, as well as possible associated issues with orbit integration accuracy, might have been present also in the previous SPH calculations.

Our findings thus suggest that, as in the case of purely stellar backgrounds, a continued decay towards the gravitational wave regime is not automatically achieved in a gaseous background.
However, concluding that there is a last parsec problem in gaseous backgrounds is highly premature. First of all, while the strength of our models, relative to other studies adopting idealized nuclear disks, relies in the realistic disk conditions inherited by the galaxy merger, there are still several important simplifications and omissions in the physics at play.
First, as in \citet{MayerSci2007} and \citet{MayerKazantzidis2008}, we considered a single phase medium described by an effective EOS. In reality the nuclear disk will host a complex multi-phase ISM, with possibly a highly inhomogeneous density structure \citep[e.g.][]{WadaNorman2001}.
Star formation and supernovae feedback will provide both diffuse and localized heating sources, and if the black holes are active as AGNs during one or more phases of the merger they would likely change the initial conditions of the nuclear disk arising after the merger (i.e. change its density and thermal structure, both being crucial for dynamical friction).
Ongoing work with a new multi-phase ISM, star formation and feedback scheme, for the moment implemented only in \GASOLINE, will soon provide a clue on the importance of such complexity in the sinking rate of the SMBH binary (Roskar et al., in preparation).
Furthermore, allowing for star formation to happen in the nuclear disk, another missing ingredient, will have an impact since stars can aid the decay in regions where the gas becomes inefficient, as long as stars move on sufficiently non circular orbit with respect to the frame of the binary, in order to keep the loss cone continuously filled \citep{Preto2010, Khan2011}.
Indeed in a massive self-gravitating disk such as the one obtained here, stars would likely exhibit centrophilic orbits as a result of non-axisymmetric instabilities. One can image thus a multi-stage decay, in which gas is the leading drag source down to parsec scales, owing to its very efficient action demonstrated here \citep{MayerSci2007}, and stars might take over at smaller separations.
Future calculations capable of capturing both a realistic multi-phase gaseous medium and the full stellar dynamical response including three-body interactions between the binary and the stars at small scales mark the necessary next frontier that will be needed to progress further in understanding the shrinking of SMBH binaries well below parsec scales.


\section*{Acknowledgments}
This work was granted access to the HPC resources of C.C.R.T. under the allocation 2010(1)-SAP2191 made by GENCI (Grand \'Equipement National de Calcul Intensif). The authors are grateful to P. Madau for fruitful discussions, D.C. acknowledges support from the Astrosim program of the European Science Foundation (grant 2009-2277) and L.M. acknowledges support from a grant of the Swiss National Science Foundation (SNSF).


\bibliography{SMBHs,romain}

\label{lastpage}
\end{document}